\pdfoutput=1
\documentclass[english,aps,prl,showpacs,twocolumn,superscriptaddress]{revtex4-1}
\usepackage{graphicx}
\usepackage{color}
\usepackage{enumerate}
\usepackage{amsmath}
\usepackage{amssymb}
\usepackage{stmaryrd}
\usepackage{multirow}
\usepackage{tikz}
\usepackage{float}
\usepackage{subcaption}
\usepackage{breqn}
\usepackage{booktabs}
\usepackage{hyperref}
\usepackage{url}
\usepackage{soul}
\def\beq{\begin{equation}}
\def\eeq{\end{equation}}	 

\usepackage{babel}
\begin{document}

\title{Finite temperature properties of strongly correlated systems\\ via
variational Monte Carlo}
\begin{abstract}
Variational methods are a common approach for computing properties of ground states but have not yet found analogous success in finite temperature calculations. In this work we develop a new variational finite temperature algorithm (VAFT) which combines ideas from minimally entangled typical thermal states (METTS), variational Monte Carlo (VMC) optimization and path integral Monte Carlo (PIMC). This allows us to define an implicit variational density matrix to estimate finite temperature properties in two and three dimensions. We benchmark the algorithm on the bipartite Heisenberg model and compare to exact results.
\end{abstract}
\author{Jahan Claes}
\author{Bryan K. Clark}
\affiliation{Department of Physics$,$ University of Illinois at Urbana-Champaign}
\maketitle

Strongly correlated fermionic and frustrated spin systems span many interesting physical systems.  Computing properties of these systems is difficult, particularly in dimensions greater then one where density matrix renormalization group (DMRG) \cite{DMRG1} methods are not particularly effective, and the fermion sign problem  \cite{SIGN} renders exact quantum Monte Carlo methods such as Path Integral Monte Carlo (PIMC) \cite{PIMC1,PIMC2} exponential. However, for ground state properties, there has been significant progress using variational Monte Carlo (VMC) methods which, while approximate, often give qualitatively and quantitatively detailed information \cite{VMC1,VMC2}. In VMC, one optimizes for the lowest energy state over a set of variational wave-functions $|\Psi[\vec{\alpha}]\rangle$ parameterized  by $\vec{\alpha}$. The approximation improves as the set of variational wave-functions expands to include wavefunctions that have greater overlap with the true ground state $|\Psi_{0}\rangle$, becoming exact in the limit where $|\Psi_{0}\rangle$ is included in the space. There exist various forms of variational wave-functions including the Slater-Jastrow \cite{Jastrow}, backflow \cite{backflow1,backflow2}, Huse-Elser \cite{HuseElser} (equivalently correlator product states \cite{CPS}, entanglement plaquette states \cite{EntanglementPlaquette}, or graph tensor network states \cite{graphTensorNetwork}), BDG states, projected entangled pair states \cite{PEPS,PEPS2}, etc. At finite temperature, variational techniques have been less useful, except again in one-dimension where minimally entangled typical thermal states (METTS) \cite{METTS1, METTS2} and finite-temperature DMRG \cite{finiteTDMRG} have proved valuable.

Of course, it is also possible to use the variational technique at finite temperature.   Most naturally one might parameterize the finite temperature many body density matrix (FTDM) $\hat{\rho}\equiv \exp[-\beta \hat{H}]/\text{Tr}(\exp[-\beta \hat{H}])$ as $\hat{\rho}[\vec{\alpha}]$, optimizing again over the parameters $\vec{\alpha}$.  However, this requires specifying a set of FTDM over which to optimize, a task that is naturally more difficult than specifying a set of wavefunctions. Additionally, this optimization needs to minimize the free energy of the system, a quantity which is harder to evaluate, requiring techniques such as coupling constant integration. For one example of this approach, see ref.~\cite{variationalDensity}.

Producing a good variational ansatz for very high or low temperature density matrices is straightforward.  At high temperature (small $\beta$), one can approximate $\exp[-\beta \hat{H}]$ by either Taylor expanding as $(1-\beta \hat{H})$ or using the Trotter break-up. At low temperature, one can instead expand in terms of the low-energy excitations, writing $\hat{\rho}=\sum_{i=1}^k \exp[-\beta E_i]|\Psi_i\rangle\langle \Psi_i|$ where $|\Psi_i\rangle$ and $E_i$ are variational estimates for the $i$th eigenvector and eigenvalue respectively and $k$ is some cutoff.  Unfortunately, this can be difficult, requiring orthogonal variational estimates of a number of excited states, and as one goes to higher temperature the number of eigenstates increases exponentially.  Nonetheless, this framework is often applied in the context of density functional theory (DFT) to produce finite temperature functionals \cite{DFT}.

At any temperature, one can approximate the density matrix as a purification $\hat{\rho} \approx \sum_i \lambda_i|\Psi_i\rangle\langle\Psi_i|$ over a set $|\Psi_i\rangle$ of variational wavefunctions. However, one cannot efficiently enumerate more than a polynomial number of $|\Psi_i\rangle$, nor is it obvious how to optimize over these wavefunctions.

In spite of these difficulties, it would be useful if we could use known variational wave-functions to compute finite temperature properties. In this paper, we describe a new variational finite temperature approach (VAFT), in which we implicitly define a variational finite temperature density matrix. Our variational FTDM is a purification of an exponentially large number of variational wavefunctions. We compute properties of this FTDM via an algorithm which stochastically samples from its diagonal without ever needing to explicitly represent it. The set of variational wavefunctions need not accurately represent highly excited states, instead needing only to accurately represent imaginary time propagation of many-body basis elements.

VAFT combines ideas from METTS, VMC, and PIMC to produce this stochastic sample. Our algorithm can be viewed either as an approximate generalization of METTS to wave-functions beyond matrix product states (MPS) or a modification of two-bead PIMC which replaces the stochastic evolution over imaginary time with an approximate deterministic one.  In fact, we will see that, in this limit, METTS and PIMC are very similar. 

The overall outline of this paper is as follows. We first describe an idealized purification of the FTDM. We then introduce a variational approximation to the FTDM which is a sum over an exponential number of variational wavefunctions, and explain how to efficiently sample from the diagonal of our variational FTDM and calculate finite temperature observables. Finally, we test the VAFT algorithm using Huse-Elser variational states on the Heisenberg model, and compare to exact results. For completeness, we describe all the necessary steps for our algorithm even when they have partial overlap with methods that already exist in the literature.

\textbf{\emph{Algorithm: }}In what follows, $\{|c\rangle\}$ is some many-body basis for our Hilbert space, and $\hat{U}$ is a unitary operator that satisfies $[\hat{U},\hat{H}]=0$. Most commonly, $\{|c\rangle\}$ is a basis of product states. $\hat{U}$ may be thought of as a change of basis that respects the symmetries of the Hamiltonian. For simplicity, the reader may take $\hat{U}=\mathbf{1}$, although we will see that other choices of $\hat{U}$ can be used to mitigate various technical issues.
-
We want to represent $\hat{\rho}$ as a purification over a set of wavefunctions. One way to do this is to write
\begingroup
\addtolength{\jot}{2pt}
$$
\begin{aligned}
\hat{\rho}&=\frac{e^{-\beta \hat{H}}}{\text{Tr}(e^{-\beta \hat{H}})}\\
&= \sum_c e^{-\beta \hat{H}/2}\hat{U}|c\rangle\langle c|\hat{U}^{\dagger} e^{-\beta \hat{H}/2}\frac{1}{\text{Tr}(e^{-\beta \hat{H}})} \\
&= \sum_c \frac{e^{-\beta \hat{H}/2}\hat{U}|c\rangle\langle c|\hat{U}^{\dagger}e^{-\beta \hat{H}/2}}{\langle c|e^{-\beta \hat{H}}|c\rangle}\frac{\langle c|e^{-\beta \hat{H}}|c\rangle}{\text{Tr}(e^{-\beta \hat{H}})} \\
&= \sum_c \frac{|\tilde{\Psi}[\beta/2;c]\rangle\langle \tilde{\Psi}[\beta/2;c]|}{\langle\tilde{\Psi}[\beta/2;c]|\tilde{\Psi}[\beta/2;c]\rangle}\tilde{p}(c)\\
\end{aligned}
$$
\endgroup
where in the last line we let $\tilde{p}(c)=\langle c|\hat{\rho}|c\rangle$ be the diagonal of the FTDM, and $|\tilde{\Psi}[\beta/2;c]\rangle$ be the (unnormalized) wavefunction $\exp[-\beta \hat{H}/2]\hat{U}|c\rangle$.

Unfortunately, there is no obvious polynomial algorithm to compute either $|\tilde{\Psi}[\beta/2;c]\rangle$ or $\tilde{p}(c)$. Instead, in this paper we work with approximate versions of these quantities $|\Psi[\beta/2;c]\rangle$ and probability distribution $p(c).$

We define our approximation $|\Psi[\beta/2;c]\rangle$ as
$$|\Psi[\beta/2;c]\rangle \equiv [\hat{P}(1-\tau \hat{H})]^{\beta/(2\tau)} \hat{U}|c\rangle$$
where the operator $\hat{P}$ projects the wavefuction into a variational space defined by the set of wave-functions $|\Psi[\vec{\alpha}]\rangle$.
Notice that $|\Psi[\beta/2;c]\rangle \approx |\tilde{\Psi}[\beta/2;c]\rangle$ and approaches it exactly in the limit of a large enough variational space and small enough $\tau.$

We can compute $|\Psi[\beta/2;c]\rangle$ via (projected) imaginary time propagation by applying the operator $\hat{P}(1-\tau \hat{H})$ repeatedly, $\beta/(2\tau)$ times.  At each application, we select a new wave-function from the variational space by selecting parameters $\vec{\alpha}'$ such that the overlap
\begin{equation}
\label{maximumoverlap}
\frac{\langle\Psi[\vec{\alpha}]|(1-\tau H)|\Psi[\vec{\alpha}']\rangle}{\sqrt{\langle\Psi[\vec{\alpha}']|\Psi[\vec{\alpha}']\rangle}}
\end{equation}
 is maximized. Eqn \ref{maximumoverlap} is maximized by choosing $\vec{\alpha}'$
such that
\begin{equation}
\label{alphaprime}
\vec{\alpha}'=\vec{\alpha}-\tau \mathbf{S}^{-1}\vec{h}
\end{equation}
where, defining $|\Psi_i\rangle\equiv \frac{\partial}{\partial \alpha_i}|\Psi\rangle$, we have
\begin{equation}
\label{hdefinition}
h_i = \frac{\langle \Psi|\hat{H}|\Psi_i\rangle}{\langle\Psi|\Psi\rangle} -\frac{\langle\Psi|\hat{H}|\Psi\rangle}{\langle\Psi|\Psi\rangle}\frac{\langle\Psi|\Psi_i\rangle}{\langle\Psi|\Psi\rangle}
\end{equation}
\begin{equation}
\label{sdefinition}
(\mathbf{S})_{ij}=\frac{\langle\Psi_{j}|\Psi_{i}\rangle}{\langle\Psi|\Psi\rangle}-\frac{\langle\Psi_j|\Psi\rangle}{\langle\Psi|\Psi\rangle}\frac{\langle\Psi|\Psi_i\rangle}{\langle\Psi|\Psi\rangle}
\end{equation}
Each of these quantities can be evaluated via Monte Carlo (see Appendix).  This process is identical to the one used in the stochastic reconfiguration optimization method \cite{stochasticReconfig}. 

We implicitly define our approximate probability $p(c)$ as the stationary distribution of a Markov chain whose state space is the many-body basis elements $|c\rangle$ and whose transition rule is defined by 
$$\textrm{Pr}(c \rightarrow c') \propto |\langle \Psi[\beta/2;c]|c'\rangle|^2$$

We can now write a concrete algorithm to do a walk over this Markov chain.
To take a Markov step from $|c\rangle \rightarrow |c'\rangle$, we
\begin{itemize}
\item Write $\hat{U}|c\rangle$ as $|\Psi[\vec{\alpha}_0]\rangle$.
\item Use projected imaginary time propagation to compute $|\Psi[\beta/2;c]\rangle$
\item Use variational Monte Carlo to select a configuration $|c'\rangle$ from the distribution $|\langle c'|\Psi[\beta/2;c]\rangle|^2$.
\end{itemize}

Note that this Markov chain may alternately be viewed as a walk over the space of wavefunctions $|\Psi[\beta/2;c]\rangle$ where $\textrm{Pr}(|\Psi[\beta/2;c]\rangle \rightarrow |\Psi[\beta/2;c']\rangle)\propto |\langle \Psi[\beta/2;c]|c'\rangle|^2$.
The Markov chain used in our approach, restricted to the space of MPS states and without the projection or operator $\hat{U}$, is the one used in METTS; there, though, steps on the Markov chain are implemented differently. 
Notice this is a legitimate Markov chain for any set of wave-functions as it is memoryless and therefore is guaranteed to reach its stationary distribution $p(c)$ in the long term limit. In the case where $|\Psi[\beta/2;c]\rangle$ is exactly equal to $|\tilde{\Psi}[\beta/2;c]\rangle$, this Markov chain has the stationary distribution $\tilde{p}(c)$ (see Appendix for proof). Thus, when $|\Psi[\beta/2;c]\rangle\approx|\tilde{\Psi}[\beta/2;c]\rangle$, we expect $p(c)\approx \tilde{p}(c)$. 

It is worth stepping back a moment and recognizing what we've achieved. Starting with some variational space defined by $|\Psi[\vec{\alpha}]\rangle$, we've implicitly defined an approximate FTDM that is a purification over an exponentially large number of variational wavefunctions.
$$
\hat{\rho} = \sum_{c} \frac{|\Psi[\beta/2;c]\rangle\langle\Psi[\beta/2;c]|}{\langle\Psi[\beta/2;c]|\Psi[\beta/2;c]\rangle}p(c)
$$
Note that this is still a valid density matrix, in that $\hat{\rho}$ is Hermitian and satisfies $\text{Tr}(\hat{\rho})=1$. In the limit where our space includes all wave-functions of the form $\exp[-\tau \hat{H}/2]\hat{U}|c\rangle$ for all $|c\rangle$ in the basis and all $0\leq\tau\leq\beta$, this reduces to the exact FTDM, in the same spirit that variational Monte Carlo becomes exact when the space includes the true ground state.

This purification is too large to explicitly enumerate. However, we can nonetheless compute the wavefunctions $|\Psi[\beta/2;c]\rangle$ as well as sample from $p(c)$ by doing a random walk over the Markov chain described above; as we'll see below, this is all that is necessary for computing finite temperature properties of this density matrix.

Note that, unlike in ground state VMC, there is no optimization step; the approximate FTDM is naturally close to the actual FTDM, and the only adjustment that can be made to our approximation is changing the underlying variational space. Where ground state VMC takes a set of variational wavefunctions as input and optimizes to select the best one, the VAFT algorithm takes a set of variational wavefunctions as input and produces a density matrix that is a weighted linear combination of all of them.  (Nonetheless, we will see later that we can still choose the best set of wave-functions among many such sets).

Up to reasonable assumptions about mixing times, this sampling is polynomial and therefore scales well with system size. The VMC steps involved in the projected imaginary propagation, the VMC in sampling the new $|c'\rangle$, and the overall Markov chain can be readily parallelized. The VMC steps can be parallelized by distributing the VMC walkers across multiple processors, while the overall Markov chain can be run independently on multiple processors in order to obtain independent samples of $p(c)$. Since the main aspects of the algorithm are applying $\hat{P}(1-\tau \hat{H})$ to a variational wavefunction and sampling from a variational wavefunction, working at finite temperature requires minimal modifications to an existing VMC code that does optimization using the stochastic reconfiguration method.

\begin{figure}
\includegraphics[width=\columnwidth]{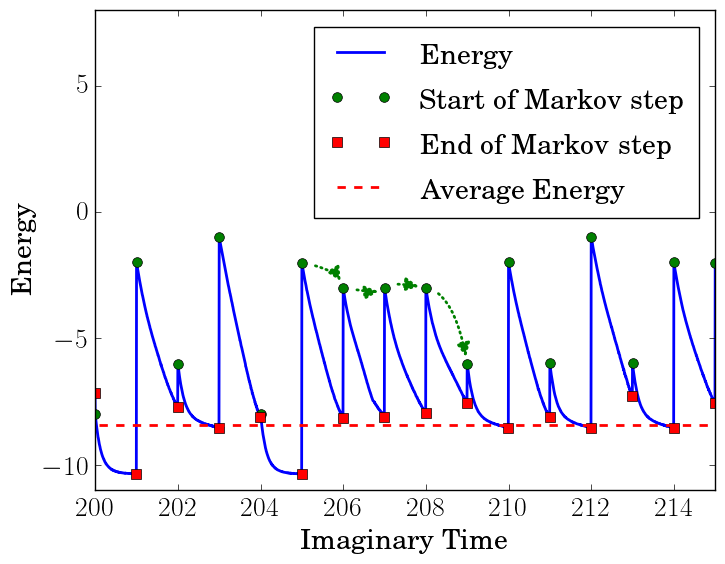}
\caption{Energy as a function of imaginary time for a $4\times 4$ square Heisenberg model, $\beta=1$, using nearest neighbor Huse-Elser states (see below). The green circles indicate the beginning of each Markov step, while the red squares indicate the end of each Markov step. The arrows show the progression of the Markov chain from one step to the next. Each Markov step lasts $\beta/2=.5$. Observables are calculated at the red circles. The dashed red line indicates the average energy of the system.}
\label{energyPlot}
\end{figure}

In fig \ref{energyPlot}, we plot the energy of the state as our algorithm runs. There are two things going on here. The overall Markov chain takes place over the basis elements $\hat{U}|c\rangle$ denoted by the green circles, or equivalently over the wavefunctions $|\Psi[\beta/2;c]\rangle$ denoted by the red squares. To get from the green circles to the red squares, we use projected imaginary time propagation (analogous to stochastic reconfiguration optimization) to apply $\hat{P}(1-\tau \hat{H})$ repeatedly. As we apply $\hat{P}(1-\tau \hat{H})$, the energy steadily decreases, until we reach the red squares, where the state has been evolved for an imaginary time of $\beta/2$. Note that while we use Monte Carlo to apply the time evolution, the time evolution is, up to statistical noise, deterministic. We calculate observables at the red squares (see below). For example, to calculate the energy, we average the energies of the wavefunctions at the red squares; this average energy is shown as the dashed line. After each imaginary time propagation, we select a new $|c\rangle$ from our wavefunction, which causes the energy to jump discontinuously to the next green circle.

There are a number of approximations in our simulation. The fundamental approximation comes from assuming some variational space of wavefunctions. There are two additional systematic approximations. First, we have introduced a time step error $\tau$ coming from approximating $\exp[-\tau \hat{H}]\approx(1-\tau \hat{H})$. Second, we have estimated the values of $(\mathbf{S})_{ij}$ and $h_j$ using VMC, so these quantities have statistical error associated with them which go away as the Monte Carlo run becomes arbitrarily long.

\textbf{\emph{Observables: }}To compute the finite temperature expectation value of an observable $\hat{\mathcal{O}}$ in our algorithm, we compute $\text{Tr}(\hat{\rho}\hat{\mathcal{O}})$ using our approximation to the FTDM:

\begingroup
\addtolength{\jot}{10pt}
$$
\langle \hat{\mathcal{O}}\rangle = \text{Tr}(\hat{\rho}\hat{\mathcal{O}})=\sum_c\langle\hat{\mathcal{O}}\rangle_c p(c)
$$
\endgroup
where
$$
\langle\hat{\mathcal{O}}\rangle_c \equiv \frac{\langle\Psi[\beta/2;c]|\hat{\mathcal{O}}|\Psi[\beta/2;c]\rangle}{\langle\Psi[\beta/2;c]|\Psi[\beta/2;c]\rangle}
$$
is simply the expectation value of $\hat{\mathcal{O}}$ in the state $|\Psi[\beta/2;c]\rangle$. Thus, to compute $\langle\hat{\mathcal{O}}\rangle$, we simply average the expectation values $\langle\hat{\mathcal{O}}\rangle_c$ across the samples $|\Psi[\beta/2;c]\rangle$ generated by our Markov chain. To compute $\langle\hat{\mathcal{O}}\rangle_c$ we use the Metropolis algorithm \cite{metropolis} to sample basis elements $|c'\rangle$ from $|\langle c'|\Psi[\beta/2;c]\rangle|^2$, then average $\langle c'|\hat{\mathcal{O}}|\Psi[\beta/2;c]\rangle/\langle c'|\Psi[\beta/2;c]\rangle$ over these samples. We can compute both diagonal and off-diagonal observables with this method; for example, in fig \ref{energyPlot}, we've computed the expectation value of the off-diagonal observable $\hat{H}$.

In the special case where $\hat{\mathcal{O}}$ is diagonal in the basis $\{|c\rangle\}$, we can also write
$$
\langle\hat{\mathcal{O}}\rangle = \sum_c \langle c |\hat{\mathcal{O}}|c\rangle p(c)
$$
Thus, rather than averaging the $\langle\hat{\mathcal{O}}\rangle_c$ over the samples, we may also estimate $\langle\hat{\mathcal{O}}\rangle$ by averaging $\langle c |\hat{\mathcal{O}}|c\rangle$ over the samples. However, this latter estimate for $\hat{\mathcal{O}}$ will generally have greater variance.


\textbf{\emph{Low Temperature Limit: }}While our algorithm is designed to compute finite-temperature properties, we can consider whether it smoothly interpolates to the ground state density matrix $|\Psi[\vec{\alpha}_g]\rangle\langle \Psi[\vec{\alpha}_g]|$ where $|\Psi[\vec{\alpha}_g]\rangle$ is the wave-function in our variational space with lowest energy.  It is straightforward to see that if $|\Psi[\vec{\alpha}_g]\rangle$ can be reached from any configuration $\hat{U}|c\rangle$ via projected imaginary time evolution, then this variational ground state limit emerges naturally. At low temperature ($\beta\rightarrow\infty)$ any initial basis configuration $\hat{U}|c\rangle$ will likely have overlap with the ground state. Thus, $\hat{U}|c\rangle$ will propagate down to the variational ground state in one application of $\exp[-\beta \hat{H}/2]$. However, it may be the case that the ground state can't be reached from all basis elements $\hat{U}|c\rangle$; instead the imaginary time evolution $\exp[-\beta \hat{H}/2]\hat{U}|c\rangle$ may get stuck in local energy minima. In such a case, our algorithm does not cleanly extrapolate to the variational limit and must have worse average energy (although it is less obvious whether other properties will be better or worse).


\textbf{\emph{Ansatz: }}The VAFT algorithm is general and can work with any variational space $\Psi[\vec{\alpha}]$, provided it includes the product states $\{\hat{U}|c\rangle$\}. The technically challenging part of the algorithm involves computing the imaginary time propagation step.  There are two technical pitfalls in this implementation that must be considered when implementing VAFT in a new variational space: linear dependence of the derivatives, and undersampling in the VMC.

A single imaginary-time step is propagated in the the subspace of derivatives $\{|\Psi_i\rangle\}$.  These derivatives may form a linearly dependent set causing $\mathbf{S}^{-1}$ to be singular.   Even when the $\{|\Psi_i\rangle\}$ are only approximately linearly dependent, the time evolution may fail; in this case, $\mathbf{S}^{-1}$ will have very large entries, which causes the parameters $\vec{\alpha}$ to change by a large amount (see eqn \ref{alphaprime}). Since the propagation is based on first-order approximations to $\delta\vec{\alpha}$, we may have to choose a very small $\tau$ for the approximation to still be valid. This can be formally resolved in various ways such as dropping parameters or using a pseudo-inverse.  It should be pointed out that in the case where we have significant linear dependence the approximate projected imaginary time evolution may be less accurate because of the lower-dimensional space. 

For most ansatz, we find this linear dependence to be most severe near a product state, $|c\rangle$, where many of the derivatives $|\Psi_i\rangle$ become identically zero. To solve this linear dependence, one can choose a $\hat{U}$ such that $\hat{U}|c\rangle$ has all nonzero derivatives.

Undersampling during the VMC generally causes the estimates of $(\mathbf{S})_{ij}$ to not converge properly. To estimate $\mathbf{S}_{ij}$, we use VMC to estimate the overlap $\langle\Psi_j|\Psi_i\rangle/\langle\Psi|\Psi\rangle$ according to (see Appendix)
$$
\frac{\langle\Psi_j|\Psi_i\rangle}{\langle\Psi|\Psi\rangle}=\frac{\sum_c  \frac{\Psi_j(c)}{\Psi(c)}\frac{\Psi_i(c)}{\Psi(c)}|\Psi(c)|^2}{\sum_c |\Psi(c)|^2}
$$
We compute this sum by sampling basis elements $|c\rangle$ from $|\Psi\rangle$. However, if there exist $|c\rangle$ such that $\Psi(c)\approx 0$, but $\Psi_j(c)\Psi_i(c) \neq 0$, our sampling will essentially ignore this $|c\rangle$, even though it is important for estimating the overlap. Thus, we undersample relevant basis elements. 

Undersampling, like linear dependence of derivatives, is most severe near a product state $|c\rangle$. In this case, every configuration $|c'\rangle\neq|c\rangle$ satisfies $\Psi(c')\approx 0$, though many of these $|c'\rangle$ are relevant for computing the overlap. There are two possible remedies to the undersampling problem. The classical approach is to sample using another probability distribution which doesn't undersample these configuration while modifying the measured properties to keep the integral the same. For example, one might sample $|\Psi\rangle+\sum_i a_i|\Psi_i\rangle$ rather than $|\Psi\rangle$. Alternatively, one can choose a $\hat{U}$ such that $\hat{U}|c\rangle$ has weight on all relevant basis elements. In both cases, we end up sampling from something far from a product state, as desired.

We will see an example below of choosing a $\hat{U}$ which solves both of these problems simultaneously for the Huse-Elser ansatz.

\emph{\textbf{Example: }}As a test, we apply our algorithm to a Heisenberg antiferromagnet on a two-dimensional square lattice, with Hamiltonian

$$
\hat{H}=\sum_{<i,j>}\vec{\hat{S}}_i\cdot\vec{\hat{S}}_j
$$
where the sum runs over all nearest neighbor sites of the lattice. We sample from the grand canonical ensemble, in which the total $S_z$ is allowed to fluctuate. The finite temperature properties of this model can be efficiently simulated by the stochastic series expansion (SSE) method; thus, it provides an ideal test for our approximate algorithm. Note that while this particular model has no sign problem, our algorithm is insensitive to the sign problem. We calculate the squared staggered magnetization, defined by
$$
\hat{\mathcal{O}} = \left(\sum_i \sigma_i\vec{\hat{S}}_i\right)^2
$$
with $\sigma_i=\pm 1$ alternating between adjacent sites. We also calculate the energy, and an estimate of the free energy.

\begin{figure}
\includegraphics[width=\columnwidth]{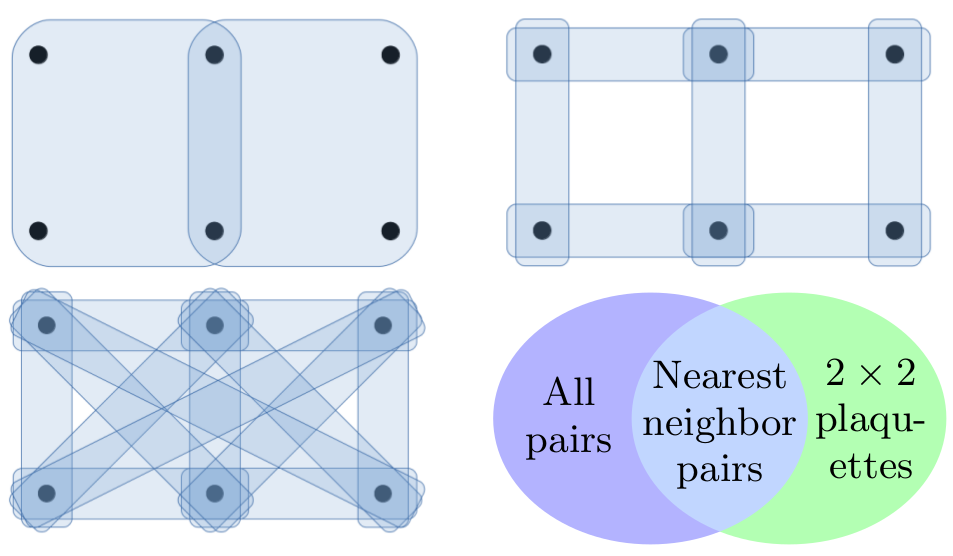}
\caption{Three possible arrangements of correlators on a lattice. Each shaded bubble represents the support of one correlator. Counterclockwise from the upper right: Nearest neighbor pairs, $2\times 2$ plaquettes, all pairs. Bottom right: The variational space defined by ``nearest neighbor pairs'' is a subset of both ``all pairs'' and ``$2\times 2$ plaquettes'', while ``all pairs'' and ``$2\times 2$ plaquettes'' are, naively, incomparable.}
\label{corrs}
\end{figure}

  For this example, we use Huse-Elser variational states \cite{HuseElser}. If we let $c_i=\pm 1$ label the spin of the $i$th particle in our lattice, the most general state $|\Psi\rangle$ can be written as $\sum_{c_1,..,c_n}\Psi^{c_1,...,c_n}|c_1,...,c_n\rangle$, where $\Psi^{c_1,...,c_n}$ is an arbitrary set of complex coefficients. In a Huse-Elser variational space, $\Psi^{c_1,...,c_n}$ is restricted to be of the form
$$
\Psi^{c_1,...,c_n}=\prod_j C_j^{c_{j_1},...,c_{j_k}}
$$
where each $C_j^{c_{j_1},...,c_{j_k}}$ is an arbitrary complex number that depends only on some subset of the lattice $\{c_{j_1},...,c_{j_k} \}$. Each $C_j$ is called a correlator, and the subset $\{c_{j_1},...,c_{j_k} \}$ is called the support of the correlator. We can specify a Huse-Elser variational space by specifying the supports of all the correlators. Three examples of possible correlator arrangements are shown in fig.~\ref{corrs}. As we increase the number of wavefunctions in our variational space, by either increasing the number of correlators or increasing the size of the correlators, we will more accurately reproduce the imaginary time evolution.

The most natural basis $\{|c\rangle\}$ for the Huse-Elser states is the basis of $\hat{S}_z$ product states, since these basis elements can be sampled in the VMC. However, in this basis, we suffer from both undersampling and linearly dependent derivatives. In fact, for Huse-Elser states, when $|\Psi\rangle=|c\rangle$, many of the $|\Psi_i\rangle$ are identically zero. We solve both of these problems by choosing $\hat{U}$ to be the operator that rotates $\hat{S}_z$ product states to $\hat{S}_x$ product states. This choice of $\hat{U}$ commutes with $\hat{H}$ since $\hat{H}$ is invariant under rotations, and it is clearly unitary.  To be able to cleanly represent the states $\hat{U}|c\rangle$ as Huse-Elser states, we add a set of single site correlators to our correlator arrangements. 

After this rotation no derivatives are zero, and $\hat{U}|\Psi\rangle$ has weight on all basis elements $|c\rangle$, so both problems are solved simultaneously. An additional effect of this rotation is to mix $S_z$ spin sectors, so we are indeed sampling from the grand canonical ensemble.

\begin{figure}
\includegraphics[width=\columnwidth]{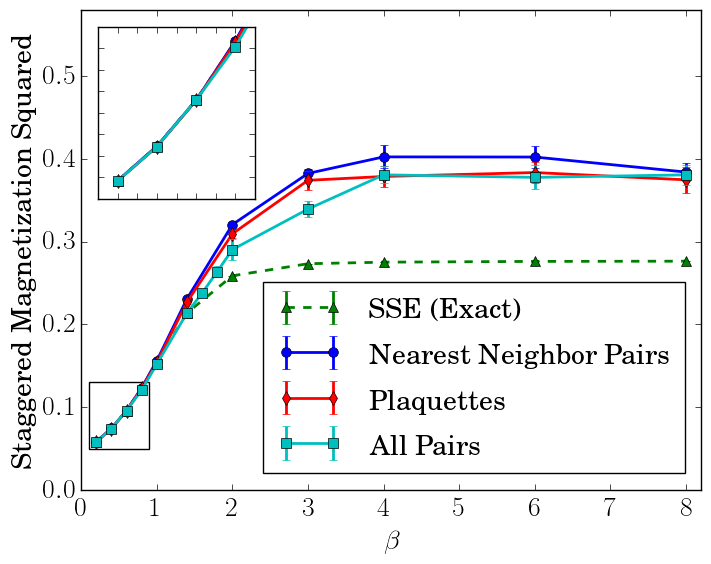}
\includegraphics[width=\columnwidth]{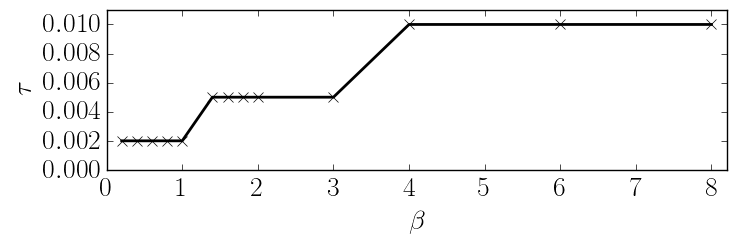}
\caption{Above: Squared staggered magnetization as a function of $\beta$ for a $4\times 4$ square Heisenberg model. Inset: Zoomed in version of the boxed region. Below: The timestep $\tau$ used for approximating $\exp[-\beta \hat{H}/2]$}
\label{fourbyfour}
\end{figure}

\begin{figure}
\includegraphics[width=1\columnwidth]{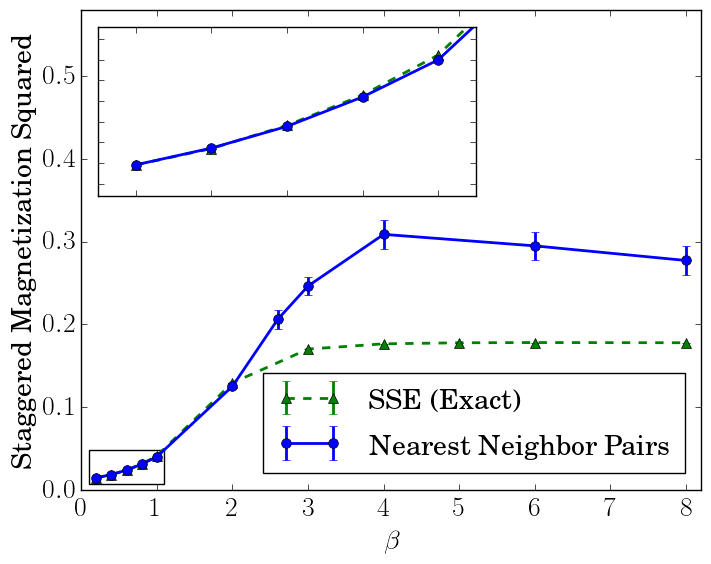}
\includegraphics[width=\columnwidth]{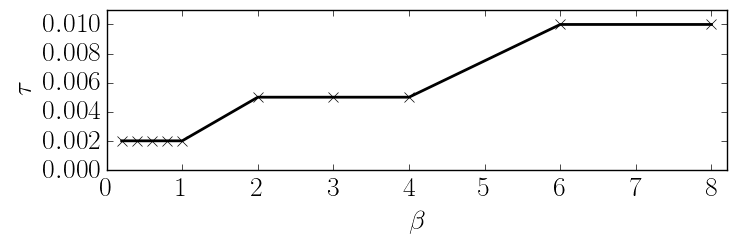}
\caption{Above: Squared staggered magnetization as a function of $\beta$ for a $8\times 8$ square Heisenberg model. Inset: Zoomed in version of the boxed region. Below: The timestep $\tau$ used for approximating $\exp[-\beta \hat{H}/2]$}
\label{eightbyeight}
\end{figure}

We plot our results in figs \ref{fourbyfour} and \ref{eightbyeight}. In fig \ref{fourbyfour}, we plot the results for three different classes of Huse-Elser states on a $4\times 4$ Heisenberg lattice, as well as the (essentially) exact result from SSE. In fig \ref{eightbyeight} we plot the results for one class of Huse-Elser states on an $8\times 8$ lattice, and the corresponding SSE results. In both cases, it is necessary to use a smaller timestep $\tau$ at lower $\beta$. It is surprising to note that despite the restriction to a variational space, for $\beta\leq 1.4$, we accurately predict the staggered magnetization, while as $\beta$ grows, our algorithm still produces a qualitatively similar curve as the true result. The divergence from the true result occurs because we can no longer represent $e^{-\tau \hat{H}}|c\rangle$ in our variational space for all $0<\tau<\beta/2$. We also see in fig \ref{fourbyfour} that as we add wavefunctions to our variational space, the variational estimate approaches the true answer.

To find the best variational space, we can compare the free energies. Each run of our algorithm gives us an approximate density matrix $\hat{\rho}(\beta)=\sum_c|\Psi[\beta/2;c]\rangle\langle\Psi[\beta/2;c]|$, where the sum runs over the samples $|c\rangle$ from the Markov chain. The ``best'' density matrix should be the one that minimizes the free energy, $F(\beta)=E(\beta)-S(\beta)/\beta$. We don't compute the entropy $S(\beta)$ directly; instead, we estimate it from the $E(\beta)$ curve via the equation
$$
S(\beta)=S(0)-\left[\int_0^\beta E(\beta')d\beta'- \beta E(\beta)\right]
$$
This will not produce the exact entropy $S(\beta)=\text{Tr}(\hat{\rho}(\beta)\ln(\hat{\rho}(\beta)))$ of our approximate density matrices, but it will produce an approximation to $S(\beta)$ based on the entire $E$ vs $\beta$ curve. We will then get an approximation to $F(\beta)$ which also depends on the whole curve.

\begin{figure}
\includegraphics[width=\columnwidth]{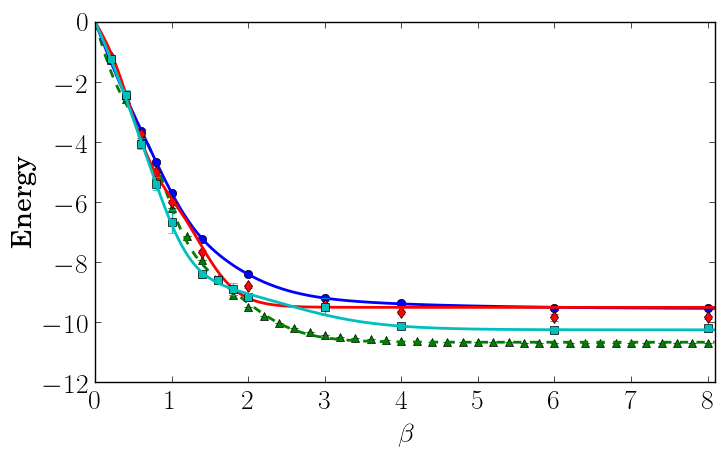}
\includegraphics[width=\columnwidth]{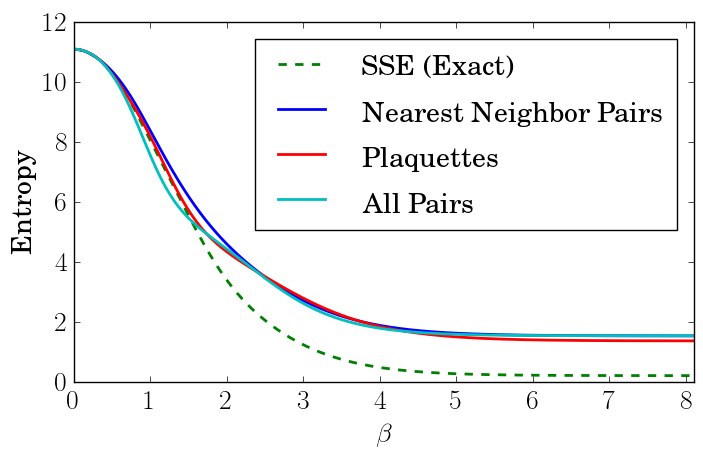}
\includegraphics[width=\columnwidth]{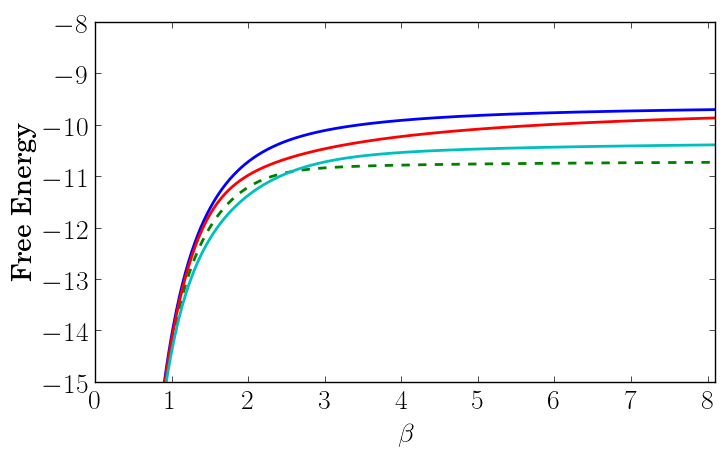}
\caption{Energy, entropy, and free energy as a function of $\beta$ for the $4\times 4$ Heisenberg lattice. The free energy is lowest for the ``all pairs'' correlator arrangement, indicating this correlator arrangement gives the best estimate for the density matrix.}
\label{free}
\end{figure}

To estimate $F(\beta)$, we fit a curve of the form
$$
E(\beta)=\frac{\sum_{i=1}^N D_i E_i e^{-E_i\beta}}{\sum_{i=1}^N D_i e^{-E_i\beta}}
$$
to our energy data, where the $E_i$ and $D_i$ are parameters representing energies and densities of states, respectively. The results of this fit for the $4\times 4$ lattice are shown in fib \ref{free}. The free energy systematically decreases as we increase the number of parameters in our variational space, indicating that ``all pairs'' gives a better density matrix than ``plaquettes'', which gives a better density matrix than ``nearest neighbor pairs.'' Note that this ordering is consistent with the fact that ``plaquettes'' and ``all-pairs'' are supersets of nearest-neighbor pairs. This approach allows us to compare our estimates, even in the absence of the exact SSE result.

Fig \ref{free} also shows that for large $\beta$, our algorithm gives us an estimate of the ground state energy. However, this estimate is strictly worse than ground state VMC. This is because not every configuration $|c\rangle$ in our Markov chain is able to propagate down to the lowest energy state in the variational space. Instead, many get stuck in local minima. Thus, the estimate for the energy at large $\beta$ is an average over local energy minima, and not the minimum energy wavefunction in our variational space.

\emph{\textbf{PIMC and METTS: }} VAFT (when $\hat{U}=\mathbf{1}$) can  either be viewed as a generalization of METTS to variational spaces besides matrix product states, or a modification of two-bead PIMC where one replaces the stochastic imaginary time evolution with a variational time evolution.

As mentioned previously, METTS uses the same Markov chain to sample from $p(c)$ as in our algorithm, with the restriction that the wavefunctions $|\Psi[\beta/2;c]\rangle$ are written as matrix product states. With this restriction, $(1-\tau\hat{H})$ can be efficiently applied using DMRG techniques rather than Monte Carlo. Thus, our algorithm may be viewed as an extension of METTS to general variational states via VMC methods.

PIMC samples from $\tilde{p}(c)$ by rewriting $\text{Tr}(\exp[-\beta \hat{H}])$ as
$$
\text{Tr}(\exp[-\beta \hat{H}])=\sum_{\{c_i\}}\prod_{i}\langle c_i|\exp[-\tau \hat{H}]|c_{i+1}\rangle
$$
where each $i$ represents a bead on the path. PIMC works by sampling paths $\{c_i\}$ with appropriate weights via the Metropolis algorithm. In the case where we have only two beads, we write
$$
\text{Tr}(\exp[-\beta \hat{H}])=\sum_{c_1 c_2}|\langle c_1|\exp[-\beta \hat{H}/2]|c_{2}\rangle|^2
$$
A local update that samples this distribution is
\begin{itemize}
\item Chose a new $|c_1\rangle$ with probability proportional to $|\langle c_1|\exp[-\beta \hat{H}/2]|c_{2}\rangle|^2$
\item Chose a new $|c_2\rangle$ with probability proportional to $|\langle c_1|\exp[-\beta \hat{H}/2]|c_{2}\rangle|^2$
\end{itemize}
This is identical to our Markov chain, provided we calculate $|\langle c_1|\exp[-\beta \hat{H}/2]|c_2\rangle|^2$ using our variational approximation to $\exp[-\beta \hat{H}/2]$; thus, VAFT may be viewed as a two-bead PIMC in which $\exp[-\beta \hat{H}/2]$ is applied variationally.

We also see that in this point of view, METTS can be viewed as a two-bead PIMC in which $\exp[-\beta \hat{H}/2]$ is applied using matrix product states.

\emph{\textbf{Discussion: }}  While we benchmarked VAFT on the bipartitie Heisenberg lattice, VAFT is general and can be applied in any setting, including continuum models, fermions, and frustrated magnets. In particular, since VAFT has no sign problem, it can be applied in cases where many other Monte Carlo methods fail. Variational Monte carlo has been the de-facto standard for understanding strongly-correlated ground states and VAFT allows this success to translate to finite temperature.  Notice, that VAFT also can work with any ansatz.

While this work has been focused on a variational approach, it is interesting to note that the ideas presented here can be used to improve calculations done in projector QMC/FCIQMC/AFQMC \cite{FCIQMC,AFQMC,AFQMC2} as well as path-integral Monte Carlo.
While the details of these approaches will be outlined in upcoming works \cite{BryanTalk}, here we mention some general aspects of the basic approach.  

With respect to projector quantum Monte Carlo, one can replace the variational imaginary time evolution in VAFT with a stochastic time evolution.  This replacement re-introduces a sign problem which then must be separately attenuated by, for example, annihilation in FCIQMC or fixed-node/phase approaches in other flavors of QMC. Such attenuation techniques have been very successful in the ground state.  Generalizing VAFT to projector QMC requires additional effort in the sampling of $|\langle c | \Psi\rangle|^2.$

This approach may also be used to improve the restricted path integral Monte Carlo algorithm (RPIMC) \cite{RPIMC1,RPIMC2}. In RPIMC, one avoids the sign problem by assuming a sign structure for the density matrix $\hat{\rho}$. However, the results of RPIMC depend on accurately guessing the sign structure. We can  compute
$\exp[-\beta \hat{H}]|R\rangle$ within our variational space for any configuration $|R\rangle$, therefore letting us evaluate the sign of the many body density matrix $\hat{\rho}(R,R';\beta)$. This technique also has positive implications for the low-temperature ergodicity problems seen in PIMC.

\textbf{Acknowledgements:} We thank Miles Stoudenmire for useful conversations.  This research is part of the Blue Waters sustained-petascale computing project, which is supported by the National Science Foundation (awards OCI-0725070 and ACI-1238993) and the state of Illinois. Blue Waters is a joint effort of the University of Illinois at Urbana-Champaign and its National Center for Supercomputing Applications. This work was also supported by SciDAC-DOE grant DE-FG02-12ER46875. We thank the Department of Energy's Institute for Nuclear Theory at the University of Washington for its hospitality where initial versions of this work were presented. Parts of this work were performed
at the Aspen Center for Physics, which is supported by
National Science Foundation grant PHY-106629.

\textbf{Appendix: }Here we prove several results stated without proof in the text. Namely, we show that the stationary distribution of the ideal Markov chain is $\tilde{p}(c)$, that eqn \ref{maximumoverlap} is maximized by choosing $\vec{\alpha}'$ according to eqn \ref{alphaprime}, and that both $h_i$ and $(\mathbf{S})_{ij}$ from eqns \ref{hdefinition} and \ref{sdefinition} can be calculated via Monte Carlo.

First, we show that the ideal Markov chain has the stationary distribution $\tilde{p}(c)$. To accomplish this, we assume that we begin in the distribution
$$
\tilde{p}(c)=\frac{\langle c|\exp[-\beta \hat{H}]|c\rangle}{\sum_{c''}\langle c''|\exp[-\beta \hat{H}]|c''\rangle}
$$
and show that a single step of the Markov chain leaves the system in the same probability distribution. In other words, we show 
$$
\sum_{c}\tilde{p}(c)\text{Pr}(c\rightarrow c')=\tilde{p}(c')
$$
where $\text{Pr}(c\rightarrow c')$ is the transition probability of our Markov chain. We first note that
\begingroup
\addtolength{\jot}{10pt}
$$
\begin{aligned}
\text{Pr}&(c\rightarrow c')\\
&=\frac{|\langle c|\hat{U}^\dagger\exp[-\beta \hat{H}/2]|c'\rangle|^2}{\sum_{c''}|\langle c|\hat{U}^\dagger\exp[-\beta \hat{H}/2]|c''\rangle|^2} \\
&=\frac{|\langle c|\hat{U}^\dagger\exp[-\beta \hat{H}/2]|c'\rangle|^2}{\sum_{c''}\langle c|\hat{U}^\dagger\exp[-\beta \hat{H}/2]|c''\rangle\langle c''|\exp[-\beta \hat{H}/2]\hat{U}|c\rangle} \\
&=\frac{|\langle c|\hat{U}^\dagger\exp[-\beta \hat{H}/2]|c'\rangle|^2}{\langle c|\hat{U}^\dagger\exp[-\beta \hat{H}]\hat{U}|c\rangle} \\
&=\frac{|\langle c|\hat{U}^\dagger\exp[-\beta \hat{H}/2]|c'\rangle|^2}{\langle c|\exp[-\beta \hat{H}]|c\rangle} \\
\end{aligned}
$$
\endgroup
Thus, we may write
\begingroup
\addtolength{\jot}{5pt}
$$
\begin{aligned}
\sum_{c}&\tilde{p}(c)\Pr(c\rightarrow c') \\
&=  \sum_c\frac{\langle c|\exp[-\beta \hat{H}]|c\rangle}{\sum_{c''}\langle c''|\exp[-\beta \hat{H}]|c''\rangle}\frac{|\langle c|\hat{U}^\dagger\exp[-\beta \hat{H}/2]|c'\rangle|^2}{\langle c|\exp[-\beta \hat{H}]|c\rangle} \\
&=\sum_c\frac{|\langle c|\hat{U}^\dagger\exp[-\beta \hat{H}/2]|c'\rangle|^2}{\sum_{c''}\langle c''|\exp[-\beta \hat{H}]|c''\rangle} \\
&=\frac{\sum_c\langle c'|\exp[-\beta \hat{H}/2]\hat{U}|c\rangle\langle c|\hat{U}^\dagger\exp[-\beta \hat{H}/2]|c'\rangle}{\sum_{c''}\langle c''|\exp[-\beta \hat{H}]|c''\rangle} \\
&=\frac{\langle c'|\exp[-\beta \hat{H}]|c'\rangle}{\sum_{c''}\langle c''|\exp[-\beta \hat{H}]|c''\rangle} \\
&=\tilde{p}(c')
\end{aligned}
$$
	\endgroup
Therefore, the stationary distribution is indeed $\tilde{p}(c)$.

Next, we prove eqn \ref{maximumoverlap} is maximized by eqn \ref{alphaprime}. Maximizing eqn \ref{maximumoverlap} is equivalent to setting the derivatives with respect to $\vec{\alpha}'$ to zero. Writing $\vec{\alpha}'=\vec{\alpha}+\delta\vec{\alpha}$, we work to first order in $\delta\vec{\alpha}$ and $\tau$. For notational convenience, we'll denote $|\Psi[\vec{\alpha}]\rangle$ by $|\Psi\rangle$, and $|\Psi[\vec{\alpha}']\rangle$ by $|\Psi'\rangle$,
\begingroup
\addtolength{\jot}{5pt}
$$
\begin{aligned}
0= & \frac{\partial}{\partial \alpha_i'}\frac{\langle\Psi|(1-\tau \hat{H})|\Psi'\rangle}{\sqrt{\langle\Psi'|\Psi'\rangle}} \\
=&\frac{\langle\Psi|(1-\tau \hat{H})|\Psi_i'\rangle}{\sqrt{\langle\Psi'|\Psi'\rangle}}-\frac{\langle\Psi|(1-\tau \hat{H})|\Psi'\rangle\langle\Psi'|\Psi_i'\rangle}{\langle\Psi'|\Psi'\rangle^{3/2}}
\end{aligned}
$$
\endgroup
Dividing both sides by $\sqrt{\langle\Psi'|\Psi'\rangle}$ gives
\begingroup
\addtolength{\jot}{5pt}
$$
\begin{aligned}
0=&\frac{\langle\Psi|(1-\tau \hat{H})|\Psi_i'\rangle}{\langle\Psi'|\Psi'\rangle} -\frac{\langle\Psi|(1-\tau \hat{H})|\Psi'\rangle}{\langle\Psi'|\Psi'\rangle}\frac{\langle\Psi'|\Psi_i'\rangle}{\langle\Psi'|\Psi'\rangle}
\end{aligned}
$$
\endgroup
Expanding the first term to first order gives
\begingroup
\addtolength{\jot}{5pt}
$$
\begin{aligned}
\frac{\langle\Psi'|\Psi_i'\rangle}{\langle\Psi'|\Psi'\rangle} -\frac{\langle\Psi_j'|\Psi_i'\rangle}{\langle\Psi'|\Psi'\rangle}\delta\alpha_j -\tau\frac{\langle\Psi|\hat{H}|\Psi_i'\rangle}{\langle\Psi'|\Psi'\rangle}
\end{aligned}
$$
\endgroup
Expanding the second term to first order gives
\begingroup
\addtolength{\jot}{5pt}
$$
\begin{aligned}
-\frac{\langle\Psi'|\Psi_i'\rangle}{\langle\Psi'|\Psi'\rangle} +\frac{\langle\Psi_j'|\Psi'\rangle}{\langle\Psi'|\Psi'\rangle}\frac{\langle\Psi'|\Psi_i'\rangle}{\langle\Psi'|\Psi'\rangle}\delta\alpha_j +\tau \frac{\langle\Psi|\hat{H}|\Psi'\rangle}{\langle\Psi'|\Psi'\rangle}\frac{\langle\Psi'|\Psi_i'\rangle}{\langle\Psi'|\Psi'\rangle}
\end{aligned}
$$
\endgroup
Thus, in total we have
\begingroup
\addtolength{\jot}{5pt}
\begin{equation}
\begin{aligned}
\label{firstorder}
0=-&\frac{\langle\Psi_j'|\Psi_i'\rangle}{\langle\Psi'|\Psi'\rangle}\delta\alpha_j -\tau\frac{\langle\Psi|\hat{H}|\Psi_i'\rangle}{\langle\Psi'|\Psi'\rangle}\\
&+\frac{\langle\Psi_j'|\Psi'\rangle}{\langle\Psi'|\Psi'\rangle}\frac{\langle\Psi'|\Psi_i'\rangle}{\langle\Psi'|\Psi'\rangle}\delta\alpha_j+\tau \frac{\langle\Psi|\hat{H}|\Psi'\rangle}{\langle\Psi'|\Psi'\rangle}\frac{\langle\Psi'|\Psi_i'\rangle}{\langle\Psi'|\Psi'\rangle}
\end{aligned}
\end{equation}
\endgroup
Since everything in eqn \ref{firstorder} is first order, we can replace $|\Psi'\rangle$ with $|\Psi\rangle$ wherever it occurs. Rearranging the equation, we get
\begingroup
\addtolength{\jot}{5pt}
\begin{equation}
\begin{aligned}
&\left(\frac{\langle\Psi_j|\Psi_i\rangle}{\langle\Psi|\Psi\rangle} +\frac{\langle\Psi_j|\Psi\rangle}{\langle\Psi|\Psi\rangle}\frac{\langle\Psi|\Psi_i\rangle}{\langle\Psi|\Psi\rangle}\right)\delta\alpha_j \\
&= -\tau\left(\frac{\langle\Psi|\hat{H}|\Psi_i\rangle}{\langle\Psi|\Psi\rangle}-\frac{\langle\Psi|\hat{H}|\Psi\rangle}{\langle\Psi|\Psi\rangle}\frac{\langle\Psi|\Psi_i\rangle}{\langle\Psi|\Psi\rangle}\right)
\end{aligned}
\end{equation}
\endgroup
Comparing this to eqn \ref{hdefinition} and eqn \ref{sdefinition}, we see this is equivalent to
\begin{equation}
(\mathbf{S})_{ij}\delta\alpha_j =-\tau h_i
\end{equation}
from which eqn \ref{alphaprime} immediately follows. We'll note that for this derivation to imply \ref{alphaprime}, $\mathbf{S}$ need not be two-sided invertible; it suffices for $\mathbf{S}^{-1}$ to be a right inverse to $\mathbf{S}$.

Finally, we show that both $h_i$ and $(\mathbf{S})_{ij}$ can be evaluated using Monte Carlo. We rewrite each term in eqns \ref{hdefinition} and \ref{sdefinition} in a form amenable to Monte Carlo calculations.
$$
\begin{aligned}
\frac{\langle \Psi|\Psi_i\rangle}{\langle\Psi|\Psi\rangle}&=\frac{\sum_c  \frac{\Psi_i(c)}{\Psi(c)}|\Psi(c)|^2}{\sum_c |\Psi(c)|^2}\\
\frac{\langle \Psi_j|\Psi_i\rangle}{\langle\Psi|\Psi\rangle}&=\frac{\sum_c  \frac{\Psi_j(c)}{\Psi(c)}\frac{\Psi_i(c)}{\Psi(c)}|\Psi(c)|^2}{\sum_c |\Psi(c)|^2}\\
\frac{\langle \Psi_i|\hat{H}|\Psi\rangle}{\langle\Psi|\Psi\rangle}&=\frac{\sum_c  \frac{\hat{H}\Psi(c)}{\Psi_i(c)}|\Psi(c)|^2}{\sum_c |\Psi(c)|^2}\\
\frac{\langle \Psi|\hat{H}|\Psi\rangle}{\langle\Psi|\Psi\rangle}&=\frac{\sum_c  \frac{\hat{H}\Psi(c)}{\Psi(c)}|\Psi(c)|^2}{\sum_c |\Psi(c)|^2}\\
\end{aligned}
$$
From these equations, it is clear that each term can be calculated by sampling the distribution $|\Psi(c)|^2$. To produce these samples, we again use the Metropolis algorithm.

\bibliography{bibliography.bib}

\begin{thebibliography}{28}%
\makeatletter
\providecommand \@ifxundefined [1]{%
 \@ifx{#1\undefined}
}%
\providecommand \@ifnum [1]{%
 \ifnum #1\expandafter \@firstoftwo
 \else \expandafter \@secondoftwo
 \fi
}%
\providecommand \@ifx [1]{%
 \ifx #1\expandafter \@firstoftwo
 \else \expandafter \@secondoftwo
 \fi
}%
\providecommand \natexlab [1]{#1}%
\providecommand \enquote  [1]{``#1''}%
\providecommand \bibnamefont  [1]{#1}%
\providecommand \bibfnamefont [1]{#1}%
\providecommand \citenamefont [1]{#1}%
\providecommand \href@noop [0]{\@secondoftwo}%
\providecommand \href [0]{\begingroup \@sanitize@url \@href}%
\providecommand \@href[1]{\@@startlink{#1}\@@href}%
\providecommand \@@href[1]{\endgroup#1\@@endlink}%
\providecommand \@sanitize@url [0]{\catcode `\\12\catcode `\$12\catcode
  `\&12\catcode `\#12\catcode `\^12\catcode `\_12\catcode `\%12\relax}%
\providecommand \@@startlink[1]{}%
\providecommand \@@endlink[0]{}%
\providecommand \url  [0]{\begingroup\@sanitize@url \@url }%
\providecommand \@url [1]{\endgroup\@href {#1}{\urlprefix }}%
\providecommand \urlprefix  [0]{URL }%
\providecommand \Eprint [0]{\href }%
\providecommand \doibase [0]{http://dx.doi.org/}%
\providecommand \selectlanguage [0]{\@gobble}%
\providecommand \bibinfo  [0]{\@secondoftwo}%
\providecommand \bibfield  [0]{\@secondoftwo}%
\providecommand \translation [1]{[#1]}%
\providecommand \BibitemOpen [0]{}%
\providecommand \bibitemStop [0]{}%
\providecommand \bibitemNoStop [0]{.\EOS\space}%
\providecommand \EOS [0]{\spacefactor3000\relax}%
\providecommand \BibitemShut  [1]{\csname bibitem#1\endcsname}%
\let\auto@bib@innerbib\@empty
\bibitem [{\citenamefont {White}(1992)}]{DMRG1}%
  \BibitemOpen
  \bibfield  {author} {\bibinfo {author} {\bibfnamefont {S.~R.}\ \bibnamefont
  {White}},\ }\href
  {http://journals.aps.org/prl/abstract/10.1103/PhysRevLett.69.2863} {\bibfield
   {journal} {\bibinfo  {journal} {Phys. Rev. Lett.}\ }\textbf {\bibinfo
  {volume} {69}},\ \bibinfo {pages} {2863} (\bibinfo {year}
  {1992})}\BibitemShut {NoStop}%
\bibitem [{\citenamefont {Loh~Jr}\ \emph {et~al.}(1990)\citenamefont {Loh~Jr},
  \citenamefont {Gubernatis}, \citenamefont {Scalettar}, \citenamefont {White},
  \citenamefont {Scalapino},\ and\ \citenamefont {Sugar}}]{SIGN}%
  \BibitemOpen
  \bibfield  {author} {\bibinfo {author} {\bibfnamefont {E.}~\bibnamefont
  {Loh~Jr}}, \bibinfo {author} {\bibfnamefont {J.}~\bibnamefont {Gubernatis}},
  \bibinfo {author} {\bibfnamefont {R.}~\bibnamefont {Scalettar}}, \bibinfo
  {author} {\bibfnamefont {S.}~\bibnamefont {White}}, \bibinfo {author}
  {\bibfnamefont {D.}~\bibnamefont {Scalapino}}, \ and\ \bibinfo {author}
  {\bibfnamefont {R.}~\bibnamefont {Sugar}},\ }\href
  {http://journals.aps.org/prb/abstract/10.1103/PhysRevB.41.9301} {\bibfield
  {journal} {\bibinfo  {journal} {Phys. Rev. B}\ }\textbf {\bibinfo {volume}
  {41}},\ \bibinfo {pages} {9301} (\bibinfo {year} {1990})}\BibitemShut
  {NoStop}%
\bibitem [{\citenamefont {Suzuki}\ \emph {et~al.}(1977)\citenamefont {Suzuki},
  \citenamefont {Miyashita},\ and\ \citenamefont {Kuroda}}]{PIMC1}%
  \BibitemOpen
  \bibfield  {author} {\bibinfo {author} {\bibfnamefont {M.}~\bibnamefont
  {Suzuki}}, \bibinfo {author} {\bibfnamefont {S.}~\bibnamefont {Miyashita}}, \
  and\ \bibinfo {author} {\bibfnamefont {A.}~\bibnamefont {Kuroda}},\ }\href
  {http://ptp.oxfordjournals.org/content/58/5/1377.short} {\bibfield  {journal}
  {\bibinfo  {journal} {Progr. Theor. Phys.}\ }\textbf {\bibinfo {volume}
  {58}},\ \bibinfo {pages} {1377} (\bibinfo {year} {1977})}\BibitemShut
  {NoStop}%
\bibitem [{\citenamefont {Ceperley}(1995)}]{PIMC2}%
  \BibitemOpen
  \bibfield  {author} {\bibinfo {author} {\bibfnamefont {D.~M.}\ \bibnamefont
  {Ceperley}},\ }\href
  {http://journals.aps.org/rmp/abstract/10.1103/RevModPhys.67.279} {\bibfield
  {journal} {\bibinfo  {journal} {Rev. Mod. Phys.}\ }\textbf {\bibinfo {volume}
  {67}},\ \bibinfo {pages} {279} (\bibinfo {year} {1995})}\BibitemShut
  {NoStop}%
\bibitem [{\citenamefont {McMillan}(1965)}]{VMC1}%
  \BibitemOpen
  \bibfield  {author} {\bibinfo {author} {\bibfnamefont {W.~L.}\ \bibnamefont
  {McMillan}},\ }\href {\doibase 10.1103/PhysRev.138.A442} {\bibfield
  {journal} {\bibinfo  {journal} {Phys. Rev.}\ }\textbf {\bibinfo {volume}
  {138}},\ \bibinfo {pages} {A442} (\bibinfo {year} {1965})}\BibitemShut
  {NoStop}%
\bibitem [{\citenamefont {Ceperley}\ \emph {et~al.}(1977)\citenamefont
  {Ceperley}, \citenamefont {Chester},\ and\ \citenamefont {Kalos}}]{VMC2}%
  \BibitemOpen
  \bibfield  {author} {\bibinfo {author} {\bibfnamefont {D.}~\bibnamefont
  {Ceperley}}, \bibinfo {author} {\bibfnamefont {G.}~\bibnamefont {Chester}}, \
  and\ \bibinfo {author} {\bibfnamefont {M.}~\bibnamefont {Kalos}},\ }\href
  {http://journals.aps.org/prb/abstract/10.1103/PhysRevB.16.3081} {\bibfield
  {journal} {\bibinfo  {journal} {Phys. Rev. B}\ }\textbf {\bibinfo {volume}
  {16}},\ \bibinfo {pages} {3081} (\bibinfo {year} {1977})}\BibitemShut
  {NoStop}%
\bibitem [{\citenamefont {Jastrow}(1955)}]{Jastrow}%
  \BibitemOpen
  \bibfield  {author} {\bibinfo {author} {\bibfnamefont {R.}~\bibnamefont
  {Jastrow}},\ }\href
  {http://journals.aps.org/pr/abstract/10.1103/PhysRev.98.1479} {\bibfield
  {journal} {\bibinfo  {journal} {Phys. Rev.}\ }\textbf {\bibinfo {volume}
  {98}},\ \bibinfo {pages} {1479} (\bibinfo {year} {1955})}\BibitemShut
  {NoStop}%
\bibitem [{\citenamefont {Feynman}\ and\ \citenamefont
  {Cohen}(1956)}]{backflow1}%
  \BibitemOpen
  \bibfield  {author} {\bibinfo {author} {\bibfnamefont {R.}~\bibnamefont
  {Feynman}}\ and\ \bibinfo {author} {\bibfnamefont {M.}~\bibnamefont
  {Cohen}},\ }\href
  {http://journals.aps.org/pr/abstract/10.1103/PhysRev.102.1189} {\bibfield
  {journal} {\bibinfo  {journal} {Phys. Rev.}\ }\textbf {\bibinfo {volume}
  {102}},\ \bibinfo {pages} {1189} (\bibinfo {year} {1956})}\BibitemShut
  {NoStop}%
\bibitem [{\citenamefont {Lee}\ \emph {et~al.}(1981)\citenamefont {Lee},
  \citenamefont {Schmidt}, \citenamefont {Kalos},\ and\ \citenamefont
  {Chester}}]{backflow2}%
  \BibitemOpen
  \bibfield  {author} {\bibinfo {author} {\bibfnamefont {M.~A.}\ \bibnamefont
  {Lee}}, \bibinfo {author} {\bibfnamefont {K.}~\bibnamefont {Schmidt}},
  \bibinfo {author} {\bibfnamefont {M.}~\bibnamefont {Kalos}}, \ and\ \bibinfo
  {author} {\bibfnamefont {G.}~\bibnamefont {Chester}},\ }\href
  {http://journals.aps.org/prl/abstract/10.1103/PhysRevLett.46.728} {\bibfield
  {journal} {\bibinfo  {journal} {Phys. Rev. Lett.}\ }\textbf {\bibinfo
  {volume} {46}},\ \bibinfo {pages} {728} (\bibinfo {year} {1981})}\BibitemShut
  {NoStop}%
\bibitem [{\citenamefont {Huse}\ and\ \citenamefont {Elser}(1988)}]{HuseElser}%
  \BibitemOpen
  \bibfield  {author} {\bibinfo {author} {\bibfnamefont {D.~A.}\ \bibnamefont
  {Huse}}\ and\ \bibinfo {author} {\bibfnamefont {V.}~\bibnamefont {Elser}},\
  }\href {http://journals.aps.org/prl/abstract/10.1103/PhysRevLett.60.2531}
  {\bibfield  {journal} {\bibinfo  {journal} {Phys. Rev. Lett.}\ }\textbf
  {\bibinfo {volume} {60}},\ \bibinfo {pages} {2531} (\bibinfo {year}
  {1988})}\BibitemShut {NoStop}%
\bibitem [{\citenamefont {Changlani}\ \emph {et~al.}(2009)\citenamefont
  {Changlani}, \citenamefont {Kinder}, \citenamefont {Umrigar},\ and\
  \citenamefont {Chan}}]{CPS}%
  \BibitemOpen
  \bibfield  {author} {\bibinfo {author} {\bibfnamefont {H.~J.}\ \bibnamefont
  {Changlani}}, \bibinfo {author} {\bibfnamefont {J.~M.}\ \bibnamefont
  {Kinder}}, \bibinfo {author} {\bibfnamefont {C.~J.}\ \bibnamefont {Umrigar}},
  \ and\ \bibinfo {author} {\bibfnamefont {G.~K.-L.}\ \bibnamefont {Chan}},\
  }\href {http://journals.aps.org/prb/abstract/10.1103/PhysRevB.80.245116}
  {\bibfield  {journal} {\bibinfo  {journal} {Phys. Rev. B}\ }\textbf {\bibinfo
  {volume} {80}},\ \bibinfo {pages} {245116} (\bibinfo {year}
  {2009})}\BibitemShut {NoStop}%
\bibitem [{\citenamefont {Mezzacapo}\ \emph {et~al.}(2009)\citenamefont
  {Mezzacapo}, \citenamefont {Schuch}, \citenamefont {Boninsegni},\ and\
  \citenamefont {Cirac}}]{EntanglementPlaquette}%
  \BibitemOpen
  \bibfield  {author} {\bibinfo {author} {\bibfnamefont {F.}~\bibnamefont
  {Mezzacapo}}, \bibinfo {author} {\bibfnamefont {N.}~\bibnamefont {Schuch}},
  \bibinfo {author} {\bibfnamefont {M.}~\bibnamefont {Boninsegni}}, \ and\
  \bibinfo {author} {\bibfnamefont {J.~I.}\ \bibnamefont {Cirac}},\ }\href
  {http://iopscience.iop.org/article/10.1088/1367-2630/11/8/083026/meta}
  {\bibfield  {journal} {\bibinfo  {journal} {New J. Phys.}\ }\textbf {\bibinfo
  {volume} {11}},\ \bibinfo {pages} {083026} (\bibinfo {year}
  {2009})}\BibitemShut {NoStop}%
\bibitem [{\citenamefont {Marti}\ \emph {et~al.}(2010)\citenamefont {Marti},
  \citenamefont {Bauer}, \citenamefont {Reiher}, \citenamefont {Troyer},\ and\
  \citenamefont {Verstraete}}]{graphTensorNetwork}%
  \BibitemOpen
  \bibfield  {author} {\bibinfo {author} {\bibfnamefont {K.~H.}\ \bibnamefont
  {Marti}}, \bibinfo {author} {\bibfnamefont {B.}~\bibnamefont {Bauer}},
  \bibinfo {author} {\bibfnamefont {M.}~\bibnamefont {Reiher}}, \bibinfo
  {author} {\bibfnamefont {M.}~\bibnamefont {Troyer}}, \ and\ \bibinfo {author}
  {\bibfnamefont {F.}~\bibnamefont {Verstraete}},\ }\href
  {http://iopscience.iop.org/article/10.1088/1367-2630/12/10/103008/meta}
  {\bibfield  {journal} {\bibinfo  {journal} {New J. Phys.}\ }\textbf {\bibinfo
  {volume} {12}},\ \bibinfo {pages} {103008} (\bibinfo {year}
  {2010})}\BibitemShut {NoStop}%
\bibitem [{\citenamefont {Verstraete}\ and\ \citenamefont
  {Cirac}(2004)}]{PEPS}%
  \BibitemOpen
  \bibfield  {author} {\bibinfo {author} {\bibfnamefont {F.}~\bibnamefont
  {Verstraete}}\ and\ \bibinfo {author} {\bibfnamefont {J.~I.}\ \bibnamefont
  {Cirac}},\ }\href {https://arxiv.org/abs/cond-mat/0407066} {\bibfield
  {journal} {\bibinfo  {journal} {arXiv:cond-mat/0407066}\ } (\bibinfo {year}
  {2004})}\BibitemShut {NoStop}%
\bibitem [{\citenamefont {Murg}\ \emph {et~al.}(2007)\citenamefont {Murg},
  \citenamefont {Verstraete},\ and\ \citenamefont {Cirac}}]{PEPS2}%
  \BibitemOpen
  \bibfield  {author} {\bibinfo {author} {\bibfnamefont {V.}~\bibnamefont
  {Murg}}, \bibinfo {author} {\bibfnamefont {F.}~\bibnamefont {Verstraete}}, \
  and\ \bibinfo {author} {\bibfnamefont {J.~I.}\ \bibnamefont {Cirac}},\ }\href
  {http://journals.aps.org/pra/abstract/10.1103/PhysRevA.75.033605} {\bibfield
  {journal} {\bibinfo  {journal} {Phys. Rev. A}\ }\textbf {\bibinfo {volume}
  {75}},\ \bibinfo {pages} {033605} (\bibinfo {year} {2007})}\BibitemShut
  {NoStop}%
\bibitem [{\citenamefont {White}(2009)}]{METTS1}%
  \BibitemOpen
  \bibfield  {author} {\bibinfo {author} {\bibfnamefont {S.~R.}\ \bibnamefont
  {White}},\ }\href
  {http://journals.aps.org/prl/abstract/10.1103/PhysRevLett.102.190601}
  {\bibfield  {journal} {\bibinfo  {journal} {Phys. Rev. Lett.}\ }\textbf
  {\bibinfo {volume} {102}},\ \bibinfo {pages} {190601} (\bibinfo {year}
  {2009})}\BibitemShut {NoStop}%
\bibitem [{\citenamefont {Stoudenmire}\ and\ \citenamefont
  {White}(2010)}]{METTS2}%
  \BibitemOpen
  \bibfield  {author} {\bibinfo {author} {\bibfnamefont {E.}~\bibnamefont
  {Stoudenmire}}\ and\ \bibinfo {author} {\bibfnamefont {S.~R.}\ \bibnamefont
  {White}},\ }\href
  {http://iopscience.iop.org/article/10.1088/1367-2630/12/5/055026/meta}
  {\bibfield  {journal} {\bibinfo  {journal} {New J. Phys.}\ }\textbf {\bibinfo
  {volume} {12}},\ \bibinfo {pages} {055026} (\bibinfo {year}
  {2010})}\BibitemShut {NoStop}%
\bibitem [{\citenamefont {Feiguin}\ and\ \citenamefont
  {White}(2005)}]{finiteTDMRG}%
  \BibitemOpen
  \bibfield  {author} {\bibinfo {author} {\bibfnamefont {A.~E.}\ \bibnamefont
  {Feiguin}}\ and\ \bibinfo {author} {\bibfnamefont {S.~R.}\ \bibnamefont
  {White}},\ }\href
  {http://journals.aps.org/prb/abstract/10.1103/PhysRevB.72.220401} {\bibfield
  {journal} {\bibinfo  {journal} {Phys. Rev. B}\ }\textbf {\bibinfo {volume}
  {72}},\ \bibinfo {pages} {220401} (\bibinfo {year} {2005})}\BibitemShut
  {NoStop}%
\bibitem [{\citenamefont {Militzer}\ and\ \citenamefont
  {Pollock}(2000)}]{variationalDensity}%
  \BibitemOpen
  \bibfield  {author} {\bibinfo {author} {\bibfnamefont {B.}~\bibnamefont
  {Militzer}}\ and\ \bibinfo {author} {\bibfnamefont {E.}~\bibnamefont
  {Pollock}},\ }\href
  {http://journals.aps.org/pre/abstract/10.1103/PhysRevE.61.3470} {\bibfield
  {journal} {\bibinfo  {journal} {Phys. Rev. E}\ }\textbf {\bibinfo {volume}
  {61}},\ \bibinfo {pages} {3470} (\bibinfo {year} {2000})}\BibitemShut
  {NoStop}%
\bibitem [{\citenamefont {Kohn}\ and\ \citenamefont {Sham}(1965)}]{DFT}%
  \BibitemOpen
  \bibfield  {author} {\bibinfo {author} {\bibfnamefont {W.}~\bibnamefont
  {Kohn}}\ and\ \bibinfo {author} {\bibfnamefont {L.~J.}\ \bibnamefont
  {Sham}},\ }\href
  {http://journals.aps.org/pr/abstract/10.1103/PhysRev.140.A1133} {\bibfield
  {journal} {\bibinfo  {journal} {Phys. Rev.}\ }\textbf {\bibinfo {volume}
  {140}},\ \bibinfo {pages} {A1133} (\bibinfo {year} {1965})}\BibitemShut
  {NoStop}%
\bibitem [{\citenamefont {Sorella}(2001)}]{stochasticReconfig}%
  \BibitemOpen
  \bibfield  {author} {\bibinfo {author} {\bibfnamefont {S.}~\bibnamefont
  {Sorella}},\ }\href
  {http://journals.aps.org/prb/abstract/10.1103/PhysRevB.64.024512} {\bibfield
  {journal} {\bibinfo  {journal} {Phys. Rev. B}\ }\textbf {\bibinfo {volume}
  {64}},\ \bibinfo {pages} {024512} (\bibinfo {year} {2001})}\BibitemShut
  {NoStop}%
\bibitem [{\citenamefont {Metropolis}\ \emph {et~al.}(1953)\citenamefont
  {Metropolis}, \citenamefont {Rosenbluth}, \citenamefont {Rosenbluth},
  \citenamefont {Teller},\ and\ \citenamefont {Teller}}]{metropolis}%
  \BibitemOpen
  \bibfield  {author} {\bibinfo {author} {\bibfnamefont {N.}~\bibnamefont
  {Metropolis}}, \bibinfo {author} {\bibfnamefont {A.~W.}\ \bibnamefont
  {Rosenbluth}}, \bibinfo {author} {\bibfnamefont {M.~N.}\ \bibnamefont
  {Rosenbluth}}, \bibinfo {author} {\bibfnamefont {A.~H.}\ \bibnamefont
  {Teller}}, \ and\ \bibinfo {author} {\bibfnamefont {E.}~\bibnamefont
  {Teller}},\ }\href {http://aip.scitation.org/doi/abs/10.1063/1.1699114}
  {\bibfield  {journal} {\bibinfo  {journal} {J. Chem. Phys.}\ }\textbf
  {\bibinfo {volume} {21}},\ \bibinfo {pages} {1087} (\bibinfo {year}
  {1953})}\BibitemShut {NoStop}%
\bibitem [{\citenamefont {Booth}\ \emph {et~al.}(2009)\citenamefont {Booth},
  \citenamefont {Thom},\ and\ \citenamefont {Alavi}}]{FCIQMC}%
  \BibitemOpen
  \bibfield  {author} {\bibinfo {author} {\bibfnamefont {G.~H.}\ \bibnamefont
  {Booth}}, \bibinfo {author} {\bibfnamefont {A.~J.}\ \bibnamefont {Thom}}, \
  and\ \bibinfo {author} {\bibfnamefont {A.}~\bibnamefont {Alavi}},\ }\href
  {http://aip.scitation.org/doi/full/10.1063/1.3193710} {\bibfield  {journal}
  {\bibinfo  {journal} {J. Chem. Phys.}\ }\textbf {\bibinfo {volume} {131}},\
  \bibinfo {pages} {054106} (\bibinfo {year} {2009})}\BibitemShut {NoStop}%
\bibitem [{\citenamefont {Blankenbecler}\ \emph {et~al.}(1981)\citenamefont
  {Blankenbecler}, \citenamefont {Scalapino},\ and\ \citenamefont
  {Sugar}}]{AFQMC}%
  \BibitemOpen
  \bibfield  {author} {\bibinfo {author} {\bibfnamefont {R.}~\bibnamefont
  {Blankenbecler}}, \bibinfo {author} {\bibfnamefont {D.~J.}\ \bibnamefont
  {Scalapino}}, \ and\ \bibinfo {author} {\bibfnamefont {R.~L.}\ \bibnamefont
  {Sugar}},\ }\href {\doibase 10.1103/PhysRevD.24.2278} {\bibfield  {journal}
  {\bibinfo  {journal} {Phys. Rev. D}\ }\textbf {\bibinfo {volume} {24}},\
  \bibinfo {pages} {2278} (\bibinfo {year} {1981})}\BibitemShut {NoStop}%
\bibitem [{\citenamefont {Sugiyama}\ and\ \citenamefont
  {Koonin}(1986)}]{AFQMC2}%
  \BibitemOpen
  \bibfield  {author} {\bibinfo {author} {\bibfnamefont {G.}~\bibnamefont
  {Sugiyama}}\ and\ \bibinfo {author} {\bibfnamefont {S.}~\bibnamefont
  {Koonin}},\ }\href
  {http://www.sciencedirect.com/science/article/pii/0003491686901077}
  {\bibfield  {journal} {\bibinfo  {journal} {Ann. Phys.}\ }\textbf {\bibinfo
  {volume} {168}},\ \bibinfo {pages} {1} (\bibinfo {year} {1986})}\BibitemShut
  {NoStop}%
\bibitem [{\citenamefont {Clark}(2013)}]{BryanTalk}%
  \BibitemOpen
  \bibfield  {author} {\bibinfo {author} {\bibfnamefont {B.}~\bibnamefont
  {Clark}},\ }\href
  {http://www.int.washington.edu/talks/WorkShops/int_13_2a/People/Clark_B/Clark.pdf}
  {\enquote {\bibinfo {title} {Algorithms for finite temperature {QMC}},}\ }
  (\bibinfo {year} {2013}),\ \bibinfo {note} {presented at Advances in quantum
  Monte Carlo techniques for non-relativistic many-body systems}\BibitemShut
  {NoStop}%
\bibitem [{\citenamefont {Ceperley}(1991)}]{RPIMC1}%
  \BibitemOpen
  \bibfield  {author} {\bibinfo {author} {\bibfnamefont {D.}~\bibnamefont
  {Ceperley}},\ }\href {http://link.springer.com/article/10.1007/BF01030009}
  {\bibfield  {journal} {\bibinfo  {journal} {J. Stat. Phys.}\ }\textbf
  {\bibinfo {volume} {63}},\ \bibinfo {pages} {1237} (\bibinfo {year}
  {1991})}\BibitemShut {NoStop}%
\bibitem [{\citenamefont {Ceperley}(1992)}]{RPIMC2}%
  \BibitemOpen
  \bibfield  {author} {\bibinfo {author} {\bibfnamefont {D.}~\bibnamefont
  {Ceperley}},\ }\href
  {http://journals.aps.org/prl/abstract/10.1103/PhysRevLett.69.331} {\bibfield
  {journal} {\bibinfo  {journal} {Phys. Rev. Lett.}\ }\textbf {\bibinfo
  {volume} {69}},\ \bibinfo {pages} {331} (\bibinfo {year} {1992})}\BibitemShut
  {NoStop}%
\end{thebibliography}%

\end{document}